\documentclass[useAMS,usenatbib]{mn2e}
\usepackage{graphicx}
\usepackage{color}
\usepackage{aas_macros}
\usepackage{amsmath}
\usepackage{here}
\newcommand{\Zsun}{{\rm Z_{\bigodot}}}
\newcommand{\Msun}{\rm M_{\bigodot}}

\defcitealias{Chiaki15}{C15}
\defcitealias{Hirano14}{H14}

\title[First galaxy SED]
      {Spectral Energy Distribution of the First Galaxies: Contribution
      from Pre-Main-Sequence Stars}

\author[H. Mitani et al.]
{Hiroto Mitani,$^{1}$\thanks{E-mail: mitani@utap.phys.s.u-tokyo.ac.jp}
Naoki Yoshida$^{1, 2, 3}, $
Kazuyuki Omukai$^{4}$, and
Takashi Hosokawa$^{5}$
\\
$^{1}$Department of Physics, The University of Tokyo, 
7-3-1 Hongo, Bunkyo, Tokyo 113-0033, Japan \\
$^{2}$Kavli Institute for the Physics and Mathematics of the Universe (WPI), 
UT Institutes for Advanced Study, \\
The University of Tokyo, Kashiwa, Chiba 277-8583, Japan \\
$^{3}$Research Center for the Early Universe, The University of Tokyo, 
7-3-1 Hongo, Bunkyo, Tokyo 113-0033, Japan\\
$^{4}$Astronomical Institute, Tohoku University, Sendai 980-8578, Japan\\
$^{5}$Department of Physics, Kyoto University, Kyoto 606-8502, Japan
}

\begin{document}

\date{}

\pagerange{\pageref{firstpage}--\pageref{lastpage}} \pubyear{2019}

\maketitle

\label{firstpage}

\begin{abstract}
  One of the major goals of next-generation space-borne and ground-based
  telescopes is to detect and characterize the first galaxies that were in place
  in the first few hundred million years after the big bang.
  We study the spectral energy distribution (SED) of the first galaxies
  and discuss the prospects for detection and identification.
  We consider very young star-forming galaxies at $z=15$ %general comment
  and incorporate the contribution from pre-main-sequence (PMS) stars.
  Unlike in the present-day galaxies, primordial protostars are
  not embedded in dusty gas clouds, and hence the light from 
  them can be visible at a wide range of wavelengths.
  We use \textsc{mesa} to follow the PMS evolution 
  and use the BT-Settl model %of Allard et al. (comment(1))
  to calculate
  the SED of individual PMS stars.
  We show that PMS stars contribute to boost the flux in the mid-infrared, and that the galaxy SED at very early evolutionary phases is overall {\it redder} than at later phases. The infrared flux contribution is comparable to that caused by emission lines powered by massive stars. 
  We argue that the contribution from PMS stars is important
  for characterizing young galaxies in the early Universe and also for target selection
  with future deep galaxy surveys.
\end{abstract}


%%%%%%%%%%%%%%%%%%%%%%%%%%%%%%%%%%%%%%%%%%%%%
% INTRODUCTION %%%%%%%%%%%%%%%%%%%%%%%%%%%%%%%%%%%
%%%%%%%%%%%%%%%%%%%%%%%%%%%%%%%%%%%%%%%%%%%%%

\section{INTRODUCTION}

Recent large observational programmes using optical/infrared and radio telescopes have 
discovered a number of galaxies near the epoch of reionization at $z=6$-$7$
(Yan et al. 2011; Dunlop et al. 2013; 
Inoue et al. 2016; Hashimoto et al. 2018),
and detected promising galaxy candidates even at $z=10$ (Oesch et al. 2018).
The distant galaxies discovered so far typically have blue
colours (e.g., Bouwens et al. 2014), suggesting a small dust content and/or
contribution from peculiar stellar populations such as massive hot stars.

The photometric and spectral features are important to understand the nature
of distant galaxies and also for efficiently detecting them.
Theoretical studies of the spectral energy distributions (SEDs) of the first galaxies (Tumlinson \& Shull 2000; Bromm, Kudritzki \& Loeb 2001; Tumlinson, Giroux \& Shull 2001; Feng et al. 2016; Barrow et al. 2018; Ceverino, Klessen \& Glover 2019) %comment(2)
suggest a few outstanding features of the first galaxies that contain Population III (Pop~III) stars.
Pop~III stars are more compact and hotter than the Pop~II
counterparts with the same mass (Ezer \& Cameron 1971), 
and hence the first galaxies 
are expected to have blue SEDs. If Pop~III stars contain an unusually
massive population (e.g., Hirano et al. 2014; 2015), strong emission lines from 
ionized hydrogen and helium are also expected (Schaerer 2003). 

Another interesting possibility especially for the first galaxies is that
there could exist a number of stars that are still at the pre-main-sequence (PMS) 
stages.
For local or low-redshift galaxies, 
the contribution from PMS stars to the total SED is often
ignored in ultraviolet to optical wavelengths,
because PMS stars are deeply embedded and hidden in dusty star-forming gas clouds.
The contribution has been considered for infrared spectra of young galaxies (Leitherer et al. 1996; Zackrisson et al. 2001).
Contrastingly, the first galaxies do not contain a substantial amount of dust. Primordial protostars are embedded in a dense primordial gas cloud
(Omukai \& Palla 2003; Yoshida et al. 2008), but
are not enshrouded with dust, and thus they may be directly visible
and can possibly give an appreciable contribution to the integrated SED
of the host galaxy.

In this {\it Letter}, we calculate the SED of the first galaxies
by including the contribution from PMS stars for the first time.
We consider cases with and without gas accretion on to PMS stars.
We place the model galaxies at $z=15$ % general comment
and study in detail the mid-infrared spectra.
Hereafter, we assume the standard lambda cold dark matter ($\Lambda$CDM) cosmology with $\Omega_{\rm m} = 1 - \Omega_{\Lambda} = 0.308$
and $H_0 = 67.8 ~{\rm km}~{\rm sec}^{-1}~{\rm Mpc}^{-1}$
(Planck Collaboration XIII 2016).

\section{Methods}
We calculate the SED of the first galaxies $S(\nu ,t)$ (flux per unit frequency at frequency $\nu$ and time $t$) by summing the contributions from main-sequence (MS) stars $F_{\mathrm{MS}}$ and from PMS stars $F_{\mathrm{PMS}}$
in a time-dependent manner as
\begin{align}
\mathrm{S}(\nu, t)&= C\Big(\int_{M_{\mathrm{min}}}^{M_1(t)}F_{\mathrm{PMS}}(\nu,t)\frac{dN}{dM}dM \\
&+ \int_{M_1(t)}^{M_{\mathrm{max}}} F_{\mathrm{MS}}(\nu,t) \frac{dN}{dM}dM \Big) \nonumber
\end{align}
with the stellar mass normalization
\begin{equation}
C^{-1}=\int_{M_{\mathrm{min}}}^{M_{\mathrm{max}}}\frac{dN}{dM}\, dM.
\end{equation}
Here, $dN/dM$ is the number of stars with mass in the range of $M \sim M + dM$, 
and $M_1(t)$ is the maximum mass of stars that 
are still in the PMS phase at $t$. Massive stars that have reached the MS 
give a fractional contribution as given by the second term in equation (1).
We also add the associated nebular continuum emission powered by the MS stars. Throughout this paper, we fix the total (PMS + MS) stellar mass of a galaxy to be $10^6 \Msun$.

We consider two variations of the stellar initial mass function (IMF)
with
\begin{equation}
\frac{dN}{dM} =
 \begin{cases}
 2C_1 M^{-1.3} &(M<0.5 M_{\odot})\\
 C_1 M^{-2.3} &(M>0.5 M_{\odot})
 \end{cases} \mbox{(Kroupa)} 
 \end{equation}
 analogous to the local IMF, and
 \begin{equation}
\frac{dN}{dM} = C_2\exp\left(-\frac{(\log M-\log M_c)^2}{2\Delta^2}\right).\ \mbox{(Log-normal)}
\end{equation}
The latter is thought to represent the mass distribution of Pop~III stars as
determined by Komiya et al. (2007) who study the origin of carbon-enhanced metal-poor stars.
We adopt their preferred values of $M_c=10\Msun$ and $\Delta=0.4$. We set the maximum mass $M_{\mathrm{max}}=100\Msun$ and the minimum mass $M_{\mathrm{\min}}=0.1\Msun$
We have also considered the Salpeter IMF.
The resulting galaxy SEDs are very 
similar to the case with Kroupa IMF presented in this {\it Letter}.%comment (4)

\subsection{PMS contribution}
We consider two independent models ;one with PMS stars with fixed masses
and the other model with accreting protostars. %comment (8)(a)
 To this end, we use the stellar evolution calculation code \textsc{mesa} version 8845 (Paxton et al. 2011% , 2013, 2015 (comment (5))
).
Since the default \textsc{mesa} code does not include a module for zero-metallicity Pop~III stars, we set the stellar metallicity to the lowest available value of $Z = 0.005 \Zsun$, and assume that a Pop~III stellar SED is well approximated by that of a $Z = 0.005 \Zsun$ star with the same mass. For accreting protostars, we do not include the contribution
from accretion luminosity. We have checked that the evolutionary tracks calculated by \textsc{mesa} 
approximately match those calculated by another numerical code of Hosokawa et al. (2013).
 For our first models without mass accretion, we consider a large number of stellar
masses with $M = 0.1, 0.2, \cdots, 1, 2,\cdots, 10, 20, \cdots, 100 \Msun$.
%comment (8)(d)
 %We adopt the fully convective star of Ushomirsky et al. (1998) as our fiducial model of the initial stellar structure.
 \textcolor{black}{With our set-up, MESA code makes an initial guess for stellar structure
 by using n=1.5 polytropes (Ushomirsky et al. 1998; Paxton et al. 2011, ).}
 
 We use the BT-Settl model to calculate the spectra of PMS stars (Allard, Homeier \& Freytag 2012). 
 In practice, we input the temperature and the surface gravity calculated by \textsc{mesa} to the BT-Settl code to obtain the spectrum of an individual PMS star. For accreting protostars,
we consider the gas accretion
rate as a primary physical quantity that determines the final stellar mass.
We assign a constant accretion rate to each star and follow the PMS evolution. 
Figure 1 shows the evolutionary tracks for PMS stars with different masses (top panel)
  and with different accretion rates (bottom panel). 
  
In order to calculate the emerging galaxy SED, we consider a group of stars with different masses
or with different accretion rates.
In the case with accretion, we set the initial mass of the protostars to be $1\Msun$, because
accreting protostars can quickly grow to be more massive than $1\Msun$, and because smaller protostars
with low accretion rates do not appreciably contribute to the total SED.
%The assigned accretion rate $\dot{M}$ varies from 0 to $ 6\times 10^{-3} \Msun \mathrm{yr}^{-1}$.  
\textcolor{black}{We assume a set of accretion rates in the range from $\dot{M}$ =0 to $6\times 10^{-3} \Msun \mathrm{yr}^{-1}$, sampled with 21 logarithmic intervals.} 
Note that the typical rate for a primordial protostar is $\dot{M} \sim 10^{-3}\Msun\mathrm{yr}^{-1}$, although
the exact accretion rate depends on a variety of physical properties of the parent cloud
and varies over an order of magnitude (e.g., Hirano et al. 2014). 
% Stars which have very high accretion rate ($\sim10^{-2}M_{\odot}\mathrm{yr}^{-1}$ can not reach MS phase. 
To each star, we assign a rate within this range
so that the resulting mass spectrum of the group of stars matches the designated IMF
(equation 3 or 4) when all the stars have landed on zero-ager main sequence (ZAMS), except that the
minimum mass is set to $1\Msun$ because of the above assumption. 
We assume that mass accretion continues until the stars reach ZAMS. 
On these assumptions, the mass is simply given as a function of time as
\begin{equation}
M(t)=
    \begin{cases}
    \dot{M}~t & (t<t_{\mathrm{ZAMS}})\\
    \dot{M}~ t_{\mathrm{ZAMS}} & (t>t_{\mathrm{ZAMS}})
    \end{cases}
\end{equation}
where $t_{\mathrm{ZAMS}}$ is the characteristic evolutionary time to reach ZAMS.
 We integrate the resulting SEDs of the PMS stars from $0.1\Msun$ to $100\Msun$ for non-accreting cases and from $1\Msun$ to $100\Msun$ for accreting cases.
\textcolor{black}{We note that, in the case of a very high accretion rate of $\sim 1~\Msun\mathrm{yr}^{-1}$,
 the dense stellar envelope can alter the emergent spectrum
 (Surace et al. 2018, 2019). Although we do not adopt such a high accretion rate,  it would be interesting to incorporate the effective reddening 
 in mid-infrared if rapidly accreting supergiant protostars in the first galaxies are considered.}% comment (8)(c) and (6)
\begin{figure}[H]
\begin{center}
  \includegraphics[width=8cm]{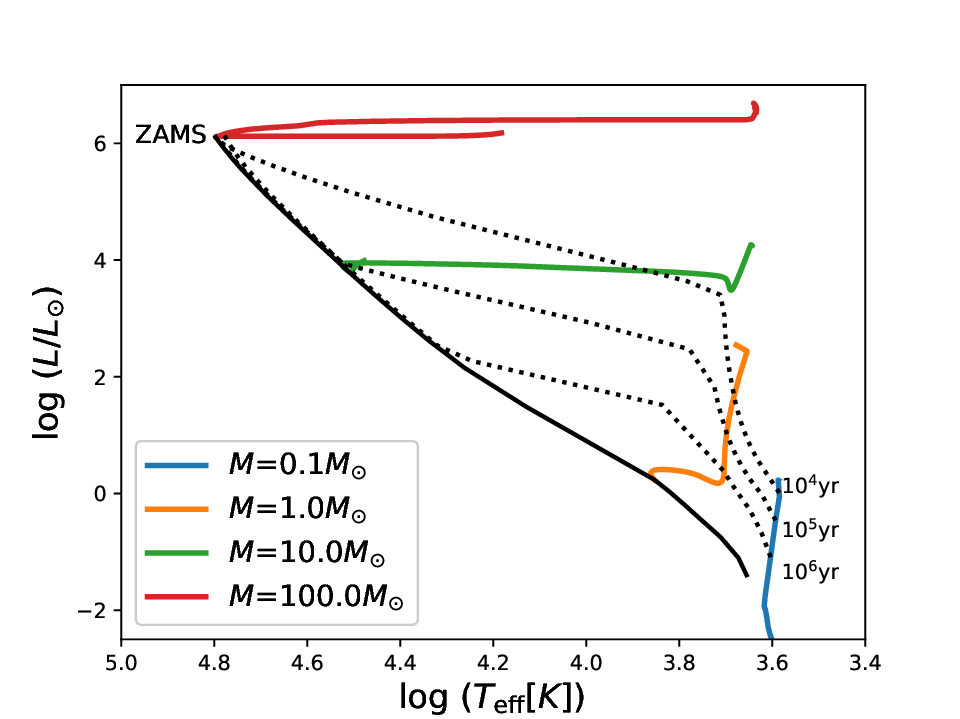}
  \includegraphics[width=8cm]{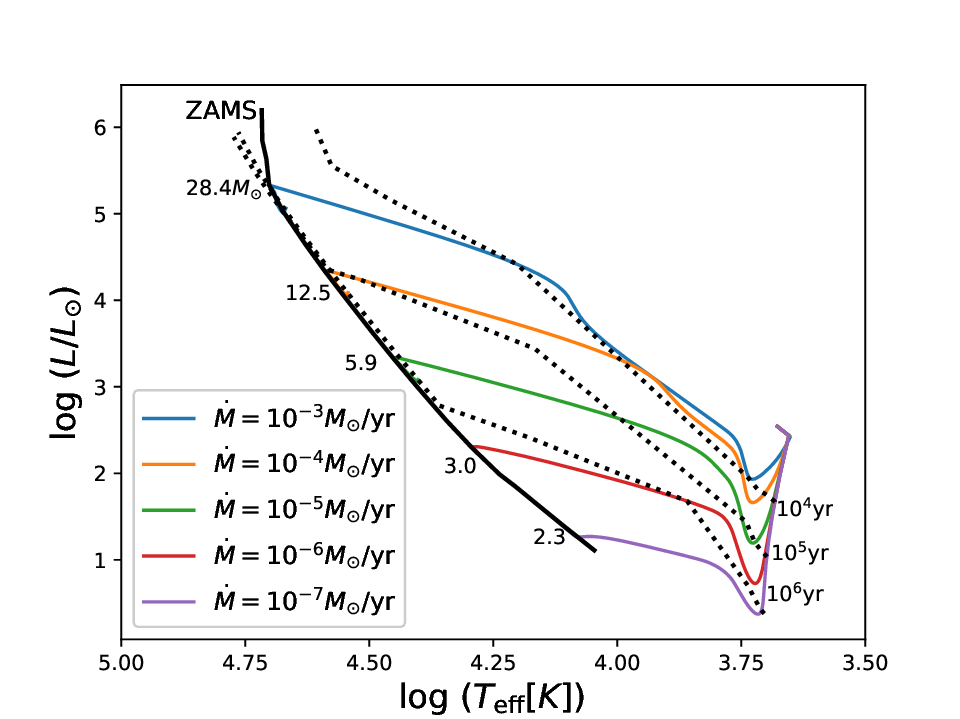}
\caption{
  PMS evolution in the Hertzsprung-Russell Diagram. Solid curves show the evolutionary tracks
  for different stellar masses (top) or for different accretion rates (bottom). The dotted curves indicate isochrones.}
\label{fig:HR}
\end{center}
\end{figure}

\subsection{MS contribution}
We use the galaxy spectral evolution code \textsc{pegase} (Fioc \& Rocca-Volmerange 1997, 2019). The nebular continuum emission is also calculated, for which we set the electron temperature of the ionized gas to be $20000 \mathrm{K}$ (e.g. Ferland 1980; Aller 1984; Osterbrock \& Ferland 2006). The MS contribution is calculated only for a limited stellar mass range $M_{1} (t)$ to $M_{\mathrm{max}}$
for the following reason.
A PMS star with mass $M_{1} (t)$ reaches the MS within a certain time $t$. When we calculate a galaxy's SED at time $t$, we include only MS stars with $M > M_1(t)$, because stars with lower masses are still in the PMS phase. For comparison, we also calculate the SED assuming all the stars are at ZAMS and show the result
in the main plots in Section 3. Since we consider very young galaxies, we do not follow the post-MS evolution.

\section{Results}
\subsection{Case with fixed-mass PMS stars}
Figure 2 shows our model galaxy SEDs at \textcolor{black}{$z=15$} % general comment
 with PMS stars at $t = 2\times10^4, 5\times 10^4,10^5$ yr after an instantaneous starburst. There is substantial PMS contribution in the mid-infrared, corresponding to rest-frame wavelength $\lambda > 0.5 \mu$m, because the PMS stars have low effective temperatures. 
Massive PMS stars give a large contribution, which  disappears, however, in $\sim 10^5$ yr because the massive stars
quickly evolve to MS (see Figure 1). 
The slight flux increase around 5 $\mu$m is owing to nebular continuum emission powered by MS stars. Other features such as the notable bumps and absorption lines are caused mainly by hydrogen
Balmer and Paschen transitions in the atmosphere of PMS stars. In the lognormal IMF case, the total flux is higher than in the Kroupa IMF case,  because there are more massive and luminous MS stars and the relative
PMS contribution is small. The rapid spectral evolution can be understood by noticing that the mid-infrared flux (in observed frame) is contributed predominantly by $\sim 10\Msun$ PMS stars that have an evolutionary time-scale of $t \sim 0.1$ Myr.
Here, we have considered cases with an instantaneous "starburst", where all the stars are born at once as PMS stars with fixed masses. We will study the case with continuous star formation 
in Section 3.3.

\begin{figure}
  \includegraphics[width=8cm]{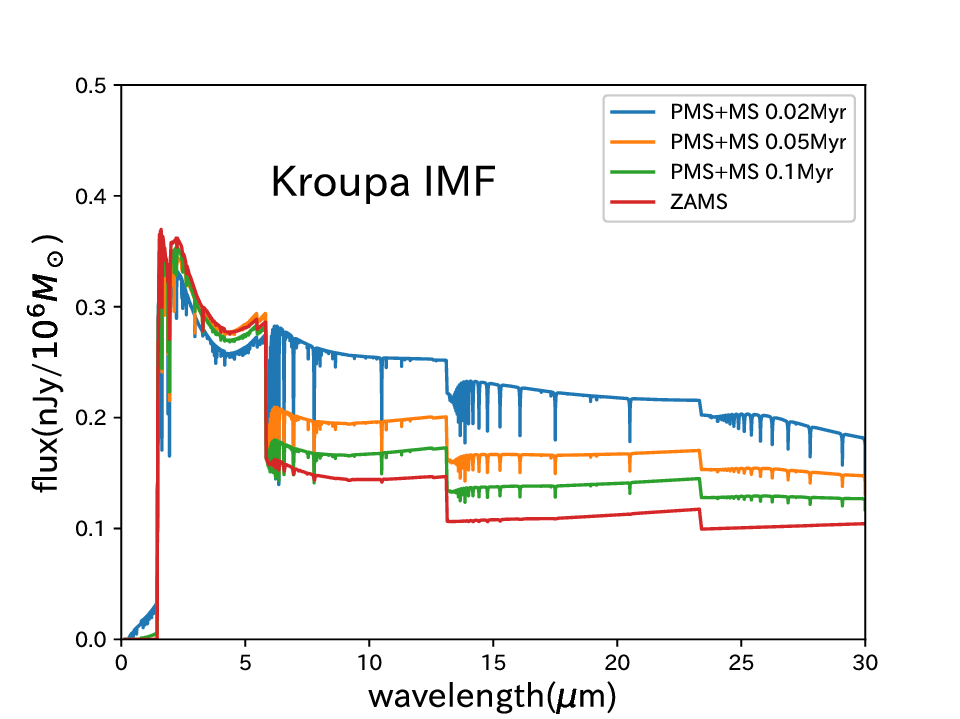}
  \includegraphics[width=8cm]{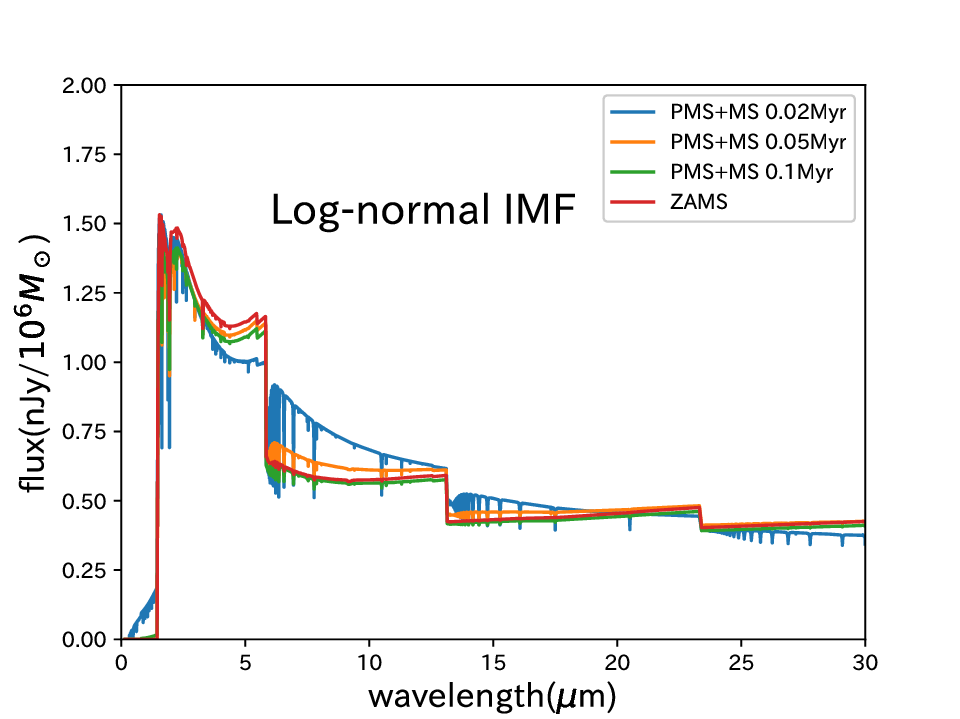}
\caption{
  SEDs of our model galaxy with Kroupa IMF (top) and Lognormal IMF (bottom) at different times. We plot the flux per unit stellar mass of $10^6\Msun$, and the galaxy
  is placed at \textcolor{black}{$z=15$}. % general comment 
  Note the mid-infrared excess contributed by PMS stars.}
\label{fig:SED1}
\end{figure}

\subsection{Case with accreting protostars}
Figure 3 shows the calculated galaxy SEDs with accreting protostars at $t = 1\times10^4, 5\times 10^4,10^5$ yr after instantaneous star formation.
The minimum stellar mass of 1 $\Msun$ in this case is greater than in the cases 
shown in Figure 2. The difference in the flux at short wavelengths (in rest-frame optical to ultraviolet bands) is largely owing to this mass difference; the accreting protostars are less massive and less luminous than when they finally reach MS (Figure 1).
Note that we plot the flux per fixed stellar mass of $10^6 \Msun$. Although the actual
stellar mass increases with time, because of mass accretion, we always
normalize the derived SED such that it corresponds to one for a $10^6 \Msun$
galaxy at the respective epoch.

\begin{figure}
  \includegraphics[width=8cm]{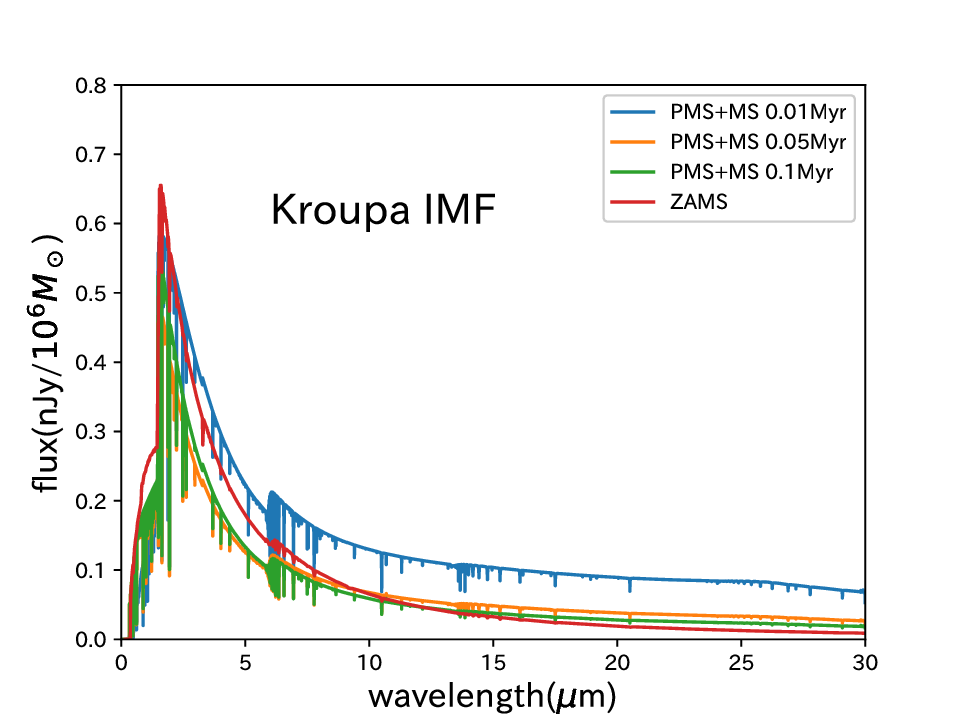}
\caption{
  SEDs of our model galaxy at \textcolor{black}{$z=15$} %general comment
   with accreting protostars at $t=1\times10^4, 5\times 10^4,10^5$ yr. We adopt the Kroupa IMF, and do not include the nebular continuum in this plot. The accreting protostars boost the mid-infrared flux, which disappears in $\sim 0.1$ Myr.}
\label{fig:SED2}
\end{figure}

\subsection{Galaxies with continuous star formation}
So far, we have considered galaxies in which star formation takes 
place instantaneously;
all the stars are born at once.
In this case, the PMS contribution to the total SED 
disappears over a typically short PMS evolutionary time. 

In real galaxies, not all the stars are formed at the same time or in the same region. 
There must always be PMS stars in a galaxy as long as star formation continues
in a star-forming region, or in many different regions that are
physically separated.
We examine a case with continuous star formation by assuming a constant rate of
10 $\Msun$ yr$^{-1}$ ($10^6 \Msun$ per 0.1 Myr). For simplicity, we consider
non-accreting PMS stars in this section.
We calculate the SED of a galaxy where $10^5\Msun$ stars are born every $10^4$ yr, 
i.e., the average star formation rate is $10\Msun$ yr$^{-1}$. 
Figure 4 show the resulting SED.
In this case, the infrared flux contribution from PMS stars is persistent. 
As long as the gas in the star-forming 
regions remains nearly primordial and contains only a small dust content, 
the newly born PMS stars can be visible. 
Metal enrichment by supernovae may quickly ruin such conditions,
but if star formation in a galaxy is sustained in a rolling manner
in many separate patches, primordial PMS
stars may contribute to the infrared flux over 
a long time-scale of galaxy formation and evolution.

\begin{figure}
  \includegraphics[width=8cm]{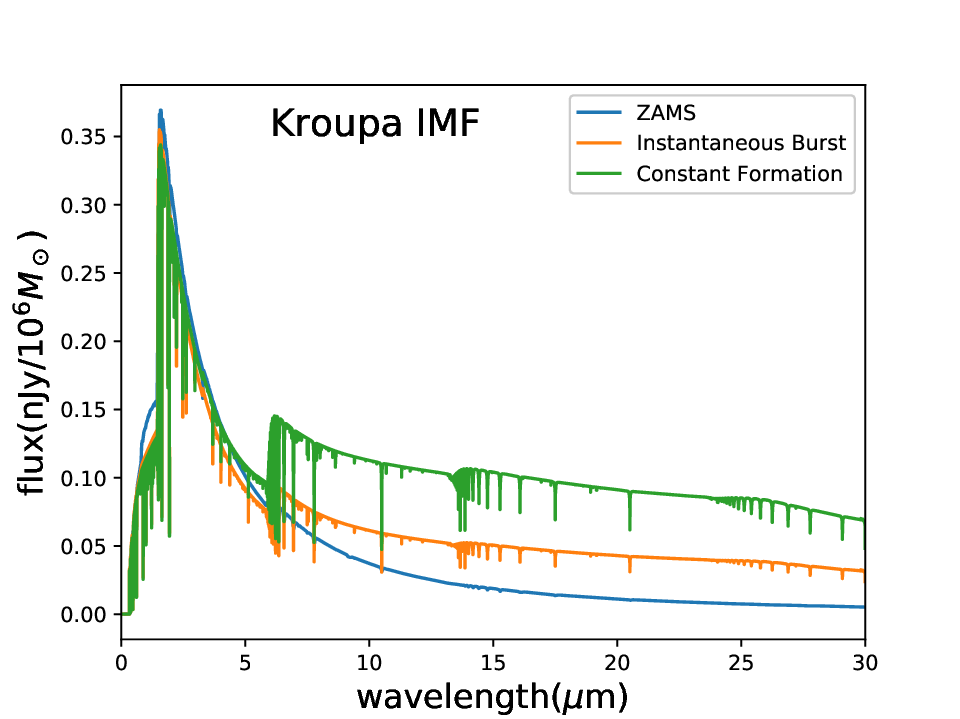}
\caption{Galaxy SED with a constant star formation rate 
of 10 $\Msun$ yr$^{-1}$ (green line).
We compare it with the SED of the corresponding case with instantaneous star formation (orange line) and with the SED of MS stars alone (blue line). 
}
\label{fig:SED_const}
\end{figure}

\textcolor{black}{\subsection{Observational feasibility}}
It is important to study the observational signatures of the PMS in the first galaxies
and to examine the feasibility of actual observations with future telescopes.
Figures 2-4 have shown consistently that the infrared flux of a $10^6 \Msun$ galaxy is below 1 nJy, suggesting that it is difficult to detect even with the 
{\it James Webb Space Telescope} (JWST) 
\footnote{https://www.jwst.nasa.gov}
\textcolor{black}{(Kalirai 2018)}. % comment (3). 
If a primordial star-forming galaxy exists at a lower redshift (e.g., Johnson 2010), and/or with help of gravitational lensing magnification, the JWST can possibly detect it, and the PMS contribution may be discerned at near-infrared bands.
A nominal detection limit of 10 nJy for the JWST NIRCam can be reached only by
a $10^7 \Msun$ galaxy at \textcolor{black}{$z=15$}. % general comment
Although this is not entirely inconceivable, 
such galaxies are likely to be already metal-enriched (see e.g., Barrow et al. 2017), and thus
the PMS stars formed and evolving in them may be dust-enshrouded. 
 
We have assumed that the light from PMS stars and protostars in the first galaxies can be directly seen. Absorption by a primordial interstellar gas or by primordial gas clouds (without dust) 
is unimportant. Also, even for rapidly accreting protostars, 
a large fraction of the stellar photosphere may be visible if the accretion occurs through a thin protostellar disc (Hosokawa et al. 2016).
The pre-cursors and discs around protostars can modify the emergent SEDs, but 
the relative contribution is usually small, 
and the main contribution in observed-frame 
infrared bands likely comes from the central stars.

\textcolor{black}{\section{Discussion and Conclusion}}
%\subsection{Line emissions}(comment (9))
It is well known that line emission owing to young stellar populations can boost the flux at the respective wavelengths (e.g., Zackrisson et al. 2001).
We calculate the emission line contribution to broad-band photometry by using \textsc{yggdrasil} (Zackrisson et al. 2011) , which adopts Pop III star models of 
Schaerer (2002). In Yggdrasil, we set the gas covering factor $f_\mathrm{cov}=0.5$ and choose the Kroupa IMF for Pop~III stars. Table 1 lists the calculated magnitude difference $\Delta m$ caused by emission lines in 10 JWST bands. 
Interestingly, the PMS contribution is comparable to, or can be even greater than, those from the line emission 
originating mainly from Balmer and Paschen transitions. Note that we here consider a primordial galaxy, and thus do not include metal lines.

For high-redshift galaxies, flux excess in near-infrared is often interpreted
as emission line contributions and/or the existence of an old stellar population
(e.g., Hashimoto et al. 2018; Tamura et al. 2019). The latter possibility is
intriguing because the existence of an old stellar population suggests
 significant star formation at an even earlier epoch in the cosmic Dark Ages.
Very young galaxies may contain a certain fraction of PMS stars
that contribute an appreciable flux in the infrared, and thus it may need to be considered
in SED modelling and fitting. A multiband observation across those 
listed in Table 1 will allow us
to distinguish the PMS contribution from that of line emission. Typically, emission lines
cause a significant boost in a few particular bands.  

\begin{table}
\centering
\begin{tabular}{lcc} \hline
Filter & $\Delta m_{\mathrm{PMS}}$ & $\Delta m_{\mathrm{line}}$ \\ \hline
%F444W & -0.43 & -0.47\\
%F560W & -0.62 & -0.53\\ 
%F770W & -0.61 & -1.35\\
%F1000W & -0.81 & -0.28\\
%F1130W & -0.79 & -0.35\\
%F1280W & -0.75 & -0.73\\
%F1500W & -0.70 & -0.35\\
%F1800W & -0.72 & -0.18\\
%F2100W & -0.59 & -0.95\\\hline
F560W & -0.04 & -0.13\\ 
F770W & -0.61 & -0.65\\
F1000W & -0.62 & -1.70\\
F1130W & -0.61 & -0.87\\
F1280W & -0.66 & -0.59\\
F1500W & -0.81 & -0.36\\
F1800W & -0.76 & -0.67\\
F2100W & -0.71 & -0.34\\
F2550W & -0.73 & -0.12\\\hline
\end{tabular}
\caption{Comparison of the flux contributions from the PMS based on the case with a Kroupa IMF in Fig. 2 in the JWST infrared bands
and those from line emissions. We list the magnitude differences. PMS stars in a very young galaxy give comparable contributions
to that from emission lines.}
\end{table}

\textcolor{black}{We have shown that PMS stars make the galaxy SED {\it redder}. Since dust extinction in the host galaxy can also redden the emergent SED, it is important to compare the relative impacts caused by PMS stars and by dust. We apply the extinction law of Calzetti (2000) with $E(B-V)=0.25$ mag to our model galaxies. We assume the same dust attenuation for both stellar and nebular components, because stars and the gas are expected to coexist in the young first galaxies. As expected, dust attenuation causes similar reddening in observed infrared bands. For two JWST bands listed in Table 1, we find $\Delta_{5-7} = m_{560} - m_{770}$ = $-0.5$ for a young galaxy
with MS stars only (see Figure 2).
PMS stars cause reddening to yield  $\Delta_{5-7} = 0.1$, whereas the corresponding colour after dust extinction is $\Delta_{5-7} = -0.2$. Interestingly, the largest difference appears in
rest-frame optical bands. The dust extinction is stronger at shorter wavelengths, whereas PMS stars boost
the flux only at longer wavelengths. 
As an illustrative example, we calculate another colour
$\Delta_{2-3} = m_{277} - m_{356}$. Adding the PMS contribution causes
little reddening in these bands, but
the dust extinction causes reddening from
$\Delta_{2-3} = -0.1$ to 0.2.
Therefore, the overall effect of PMS stars and that of dust can be,
in principle, distinguished by using multicolour diagnostics such as
$\Delta_{2-3} and \Delta_{5-7}$.}

%\subsection{The effect of the precursor and disk}
%\subsection{Conclusion} (comment (9))
We have studied the SEDs of the first galaxies that contain PMS stars. Although the time-scale of PMS evolution is short ($\sim 0.1$-1 Myr), it may be necessary to include the PMS contribution in order to model properly and interpret the SED of a young galaxy.
The PMS signature appears primarily as an excess
in the observed-frame infrared flux especially when a galaxy
sustains continuous star formation (Section 3.3).
If we do not consider the PMS contribution in the SED fitting, 
we may misestimate galaxy properties such as the total stellar mass and the star formation history.
Although our result may not readily be applicable to galaxies that are already metal- and dust-enriched, galaxies with a low dust content
likely exist in the early Universe, and PMS stars might imprint
distinct signatures similar to those presented in this {\it Letter}. 
A small mass fraction of PMS stars can contribute to the
total infrared luminosity of a galaxy. 
More detailed study is clearly warranted to understand
the photometric and spectral features of young galaxies 
in the early Universe.

\section*{acknowledgments}
NY thanks K. Shimasaku for fruitful discussion.
NY acknowledges the financial support from Japan Science and Technology Agency CREST JPMHCR1414. This work was supported by Japan Society for the Promotion of Science KAKENHI Grant number JP16H05996, JP17H06360, JP17H01102, and JP17H02869.

%%%%%%%%%%%%%%%%%%%%%%%%%%%%%%%%%%%%%%%%%%%%%%
% APPENDIX  %%%%%%%%%%%%%%%%%%%%%%%%%%%%%%%%%%%%%%
%%%%%%%%%%%%%%%%%%%%%%%%%%%%%%%%%%%%%%%%%%%%%%
%\appendix

%%%%%%%%%%%%%%%%%%%%%%%%%%%%%%%%%%%%%%%%%%%%%
% REFERENCES %%%%%%%%%%%%%%%%%%%%%%%%%%%%%%%%%%%%
%%%%%%%%%%%%%%%%%%%%%%%%%%%%%%%%%%%%%%%%%%%%%

\end{document}